# Novel properties of two-dimensional Janus transition metal hydrosulfides: electride states, charge density waves and superconductivity

Dawei Zhou[a], Dominik Szczęśniak[b,*], Pan Zhang[a], Zhuo Wang[c], Chunying Pu[*,a]

[a]College of Physics and Electronic Engineering, Nanyang Normal University, Nanyang 473061, China

[b]Institute of Physics, Jan Długosz University in Częstochowa, 13/15 Armii Krajowej Ave., Częstochowa 42200, Poland

[c]College of Intelligent Manufacturing and Electrical Engineering, Nanyang Normal University, Nanyang 473061, China

*Corresponding author.

E-mail address: d.szczesniak@ujd.edu.pl (D. Szczęśniak), puchunying@126.com ( Pu Chunying)

**Abstract**

Inspired by recent experimental synthesis of the two-dimensional Janus material MoSH, we have performed extensive first-principles calculations to investigate the characteristics of all possible Janus two-dimensional transition metal hydrosulfides (JTMSHs), in both the 2H and 1T phases. Our investigations revealed that the JTMSHs can form a unique family of two-dimensional materials with intriguing physical and chemical properties. In details, we have found that JTMSHs can exist in various states, exhibiting metallic, semiconducting, and magnetic character. One particularly intriguing finding considered identification of electride states with distinct bonding characteristics in the 2H-JTMSHs for TM=V, Nb, Ta, Mo, W and Tc. Additionally, we have observed evidence of charge density wave scenario in 1T-JTMSH with TM=Tc, Re, and W and 2H-JTMSH with TM = Tc. Importantly, by applying compressive strain to these materials, the charge density wave can be completely suppressed, eventually leading to the superconducting phase. In particular, we have shown that when subjected to a compressive strain within 10%, the superconducting transition temperature ($T_C$) of 1T-WSH, 1T-TcSH and 2H-TcSH can reach values of 13.8, 16.2 and 24.2 K respectively. Moreover, our investigation also unveiled two intrinsic phonon-mediated superconductors, namely 2H-WSH and 1T-RuSH with $T_C$ of 17.0 and 8 K, respectively.

## 1. Introduction

Two-dimensional (2D) transition metal dichalcogenides (TMDs) are considered one of the most intriguing mesoscopic materials due to their rich physics and chemistry[1-11]. Typically, 2D TMDs are composed of two types of atoms: transitional metals (TM, *i.e.* molybdenum, tungsten) and chalcogenide atoms (X, *i.e.* sulfur, selenium, tellurium). These are arranged in a graphene-like layered structure, where layer of TM atoms is encapsulated between two layers of X atoms[12]. As such, the 2D TMDs present diverse structural and electronic properties. In particular, depending on the arrangement of atoms in these materials, they exhibit distinct phases such as the octahedral 1T or trigonal prismatic 2H and 3R phases[13-15]. As a consequence, they serve as perfect low-dimensional hosts for various states, including metallic, semiconducting, insulating, and superconducting states.

In the context of the above, 2D TMDs can serve as a foundational platform for creating other 2D materials with novel functionalities. Specifically, in recent years it has been shown that chalcogenide atoms in TMDs can be substituted by using the selective epitaxial atomic replacement (SEAR) method. Such artificial replacement disrupts the mirror symmetry in the original 2D TMD materials, resulting in the formation of a new type of 2D material known as the Janus-transitional metal dichalcogenides (JTMDs). This is in addition to the already broken inverse symmetry in conventional TMDs, which is responsible for their semiconducting properties that distinguish them from the parent semimetallic graphene[16,17]. Many of JTMDs, such as the MoSSe[16,17], WSSe thin-film[16], and PtSSe[18], has been successfully synthesized. For example, Janus 2D MoSSe is formed by replacing the top-layer S(Se) of $MoS_2$($MoSe_2$) with Se(S) atoms. As a result, a new asymmetric structure can be created that allows for further fine tuning of physical and chemical properties of TMDs. This application potential of JTMDs has been already shown in terms of *e.g.* hydrogen evolution reaction process (HER)[19], water splitting[20],[21], photocatalytics[22], or the design of solar sensors[25],[23], optoelectronic nanodevices[24] and nano-piezoeletronics[25].

More interestingly, researchers further successfully stripped the top-layer S of MoS$_2$ using H atoms, which led to the formation of yet another new Janus 2D material known as 2H-MoSH[16]. For the 2H-MoSH, this material exhibits metallic properties and have been predicted to host superconducting phase with T$_C$ of 28.58 K[26]. On the other hand, its 1T phase was found to display charge density waves (CDW) and superconductivity under strain[27]. In addition to the MoSH material, our previous study suggested existence of Janus NbSH[28]. Similarly to the former, the NbSH features rich number of various phases, including 1T, 2H, and a newly predicted metallic phase. It was shown that the 1T and 2H phases are semiconducting, while the observed metal phase exhibits intrinsic superconducting behavior.

However, there is still a lack of knowledge regarding the properties of the entire Janus two-dimensional transition metal hydrosulfides (JTMSHs), except for the mentioned MoSH and NbSH materials. This raises the question of whether the JTMSHs can offer more surprises and unique properties in comparison to JTMDs. To answer that, we conduct a large-scale computer simulation by using first-principle calculations and explore the stability as well as properties of all possible 2H and 1T phases of JTMSHs. The particular attention is given to the electronic, magnetic and superconducting characteristics of the predicted structures. Furthermore, the charge density wave scenario and electride capabilities are discussed. By elucidating these characteristics, our aim is to gain deeper understanding of the JTMSH family and promote related future fundamentals studies as well as potential applications.

## 2. Computational Methods

The electronic structure and energy calculations of JTMSHs were performed by using the Vienna Ab initio Simulation Package (VASP)[29]. The projector augmented wave (PAW) method[30] was used to estimate interactions between the valence electrons and the core ions. The calculations employed the generalized gradient approximation (GGA)[31] with the Perdew-Burke-Ernzerhof (PBE) functional to estimate the exchange-correlation energy. The semi-empirical dispersion correction of DFT-D3 method was adopted to consider the Van der Waals interaction during the band and energy calculations[32]. A plane-wave cut-off energy was chosen as 600 eV to ensure accurate results.

The structure was relaxed until the residual force on each atom was smaller than 0.001 eVÅ$^{-1}$. The calculations were carried out by sampling the Brillouin zone using the 14×14×1 Monkhorst-Pack grid[33]. To eliminate interlayer interactions, a vacuum thickness of 25 Å was used. For the antiferromagnetic JTMSHs, the band structures were calculated by using the so-called band unfolding method[34].

The Allen-Dynes formula, derived from the Bardeen-Cooper-Schrieffer (BCS) theory[35], was used to estimate the T$_c$ in the following manner:

$$T_C = \frac{\omega_{\log}}{1.20} \exp\left(-\frac{1.04(1+\lambda)}{\lambda-\mu^*(1+0.62\lambda)}\right) \quad (1)$$

where $\omega_{\log}$ is the logarithmic average of the phonon energy, $\lambda$ states for the electron-phonon coupling constant (EPC) and $\mu^*$ denotes the electron-electron coulomb repulsion parameter.

The $\omega_{\log}$ and $\lambda$ parameters were calculated according to the equation (2) and equation (3), respectively[36]:

$$\omega_{\log} = \exp\left(\frac{2}{\lambda}\int_0^\infty \alpha^2 F(\omega)\log\omega \frac{d\omega}{\omega}\right) \quad (2)$$

$$\lambda = 2\int_0^\infty \frac{\alpha^2 F(\omega)}{\omega} d\omega \quad (3)$$

The electron-phonon spectral function $\alpha^2 F$ (the Eliashberg function) in the equation (3) was computed based on the phonon linewidth $\gamma$ according to equation (4):

$$\alpha^2 F(\omega) = \frac{1}{2\pi N(\varepsilon_F)} \sum_{q\nu} \delta(\omega-\omega_{q\nu}) \frac{\gamma_{q\nu}}{\hbar\omega_{q\nu}} \quad (4)$$

where $N(\varepsilon_F)$ represents the density of states (DOS) at the Fermi level, whereas $\gamma_{q\nu}$ dentoes the phonon mode's linewidth with wave vector q.

The superconducting properties were analyzed within the density functional perturbation theory, as implemented in the QUANTUM ESPRESSO package[37]. In particular, the ultrasoft pseudopotentials[38] with a kinetic cutoff energy of 45 Ry were used. The EPC constant calculations were performed by using fine electron and phonon grids (16×16×1 and 8×8×1, respectively) to allow proper interpolation. A Gaussian width of 0.02 Ry was used in the EPC calculations.

**3. Results and discussion**

**3.1 Structures and thermodynamic stability of JTMSHs**

The 2H and 1T phases are the two most common types of atoms arrangement in case of 2D TMDs[15]. As already mentioned, the 2H phase is typically a semiconducting one, while the 1T structure is usually a metastable metallic phase. By analogy, these phases are also expected to be the predominant structures in terms of the JTMSHs. Fig.1(a) illustrates the top view diagrams and side views of such 2H-JTMSH and 1T-JTMSH systems. Both phases exhibit a layered structure with one unit per cell, while also sharing the same space group, namely $P3m1$. The primary difference between the 2H and 1T phases lies in the arrangement of hydrogen (H) atoms. In particular, the 1T-JTMSH phase can be obtained by shifting the H-layers relative to the TM-layers of 2H-JTMSH.

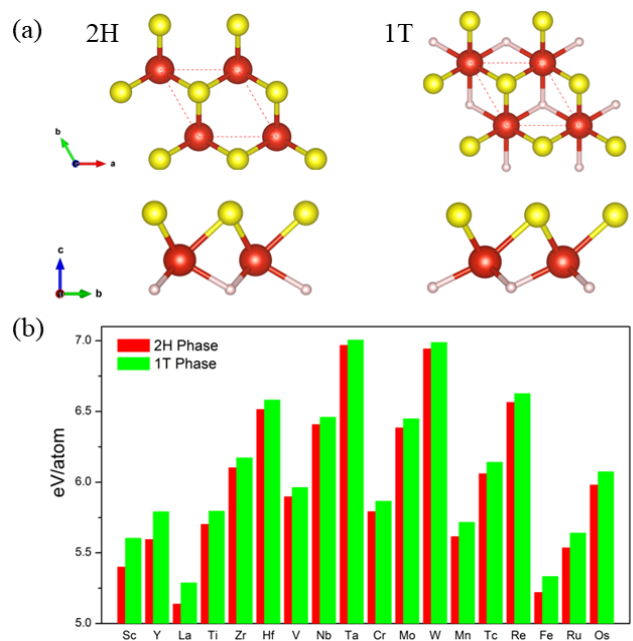

Fig.1 (a) The 2H and 1T structures of JTMSHs. The while, and yellow atoms are Hydrogen and Sulfur atoms, respectively, while the red atoms are TM atoms  (b) The cohesive energy of the 2H/1T JTMSHs.

The thermodynamic stability of the predicted structures is investigated by calculating the cohesive energy defined as:

$$E_{coh} = (E_{TM} + E_S + E_H - E_{JTMSH})/3 \tag{5}$$

where $E_{JTMSH}$, $E_{TM}$, $E_S$ and $E_H$ represent the total energies of one unit cell as well as the isolated TM, S and H atoms, respectively. Hence, higher cohesive energy signifies more stable structure, according to this definition.

The cohesive energy of the 2H and 1T phases is found to range from 5.13 eV/atom to 6.96 eV/atom and from 5.28 eV/atom to 7.0 eV/atom, respectively. In what follows, 1T-JTMSH exhibits slightly higher cohesive energy in comparison to 2H-JTMSH with the same TM element. Still for both types of JTMSHs materials the cohesive energies increase when moving down the periodic table within the same subgroup of TM elements, excluding the LaSH case. Interestingly, these values surpass most of typical 2D materials such as MoS$_2$ (4.97 eV/atom)[39], silicene (3.98eV/atom), germanene (3.26eV/atom)[40], and phosphorene (3.61eV/atom)[41], demonstrating thermodynamic stability of JTMSHs as a 2D materials. In fact, the obtained cohesive energy values of JTMSHs with TM=Hf, Ta, W, Re are similar to that of the experimentally synthesized MoSH, implying that the predicted materials have high potential for future synthesis.

**3.2 Semiconducting and metallic properties of JTMSHs**

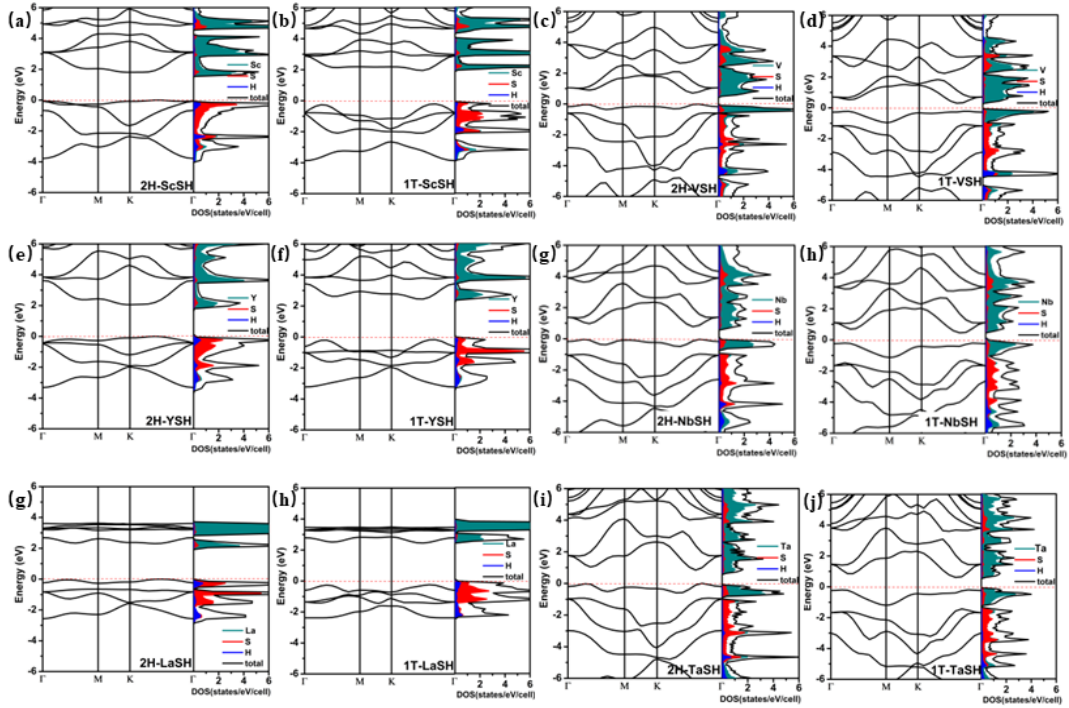

Fig.2 Band structrures and densitity of states of the non magnetic semiconducting JTMSHs.

In the present work, the calculated total density of states (DOS) of all the JTMSHs, including the spin-polarization effects, are shown in Fig.S1 in the supplementary materials. Based on these results it can be observed that JTMSHs exist in various electronic states, but predominantly in the metallic and semiconducting ones. In Fig. 2, the electronic band structures and DOSs of the semiconducting JTMSHs are presented. Interestingly, it is clear that the JTMSHs with IIIB and VB TM subgroup elements are indirect-gap semiconductors. The band gaps of 2H-JTMSHs with TM=Sc, Y, La, V, Nb, and Ta are 1.73, 1.96, 2.24, 0.45, 0.42, and 0.43 eV, respectively. On the other hand, the corresponding band gaps of 1T-JTMSHs are 2.16, 2.60, 2.62, 0.49, 0.64, and 0.80eV, respectively. Thus, in contrast to 2H-JTMSHs, the 1T phases of the JTMSHs with the same TM element are characterized by significantly larger band gaps.

Since the investigations of 2D JTMSHs were conducted under zero temperature and at zero pressure conditions, it is evident that the $p$-electrons of S, $d$-electrons of TM, and $s$-electrons of H primarily contribute to the formation of chemical bonds in the JTMSHs, according to the valence electron theory. An analysis of the DOSs provides further insights into the electronic structures of JTMDHs. As shown in Fig.3, it is found that for JTMSHs with IIIB TM subgroup elements, the valence band maximum (VBM) consists mainly of the *d*-electrons of TM atoms, while the conduction band minimum (CBM) is contributed by the *p*-electrons of S atoms. In contrast, for 2D JTMSHs with VB TM subgroup elements, both the VBM and CBM are primarily contributed by the *d*-electrons of TM atoms. This implies that bonding in JTMSHs with VB TM subgroup elements is mainly influenced by the TM atoms themselves, rather than the surrounding S atoms. Thus, JTMSHs with the IIIB TM subgroup elements are mainly bonded by S and TM atoms, while bonds in JTMSHs with the VB TM subgroup comes from the contribution of TM atoms.

Upon further examination of the total density of states shown in Fig.S1, it can observe that the 1T/2H JTMSHs with IIIB and VB TM subgroup elements are nonmagnetic. However, magnetic small gap semiconductors such as the 1T-JTMSHs with TM=Cr, Mn, and 2H-JTMSHs with TM=Ti, Zr, Hf, were also discovered (see band structures depicted in Fig.S2). These magnetic structures were further investigated using a 2×2 supercell. The determined magnetic ground state energy indicates that the 2H-ZrSH, 2H-HfSH and 1T-CrSH have spin-polarized antiferromagnetic (AFM) character, while the 2H-TiSH and 1T-MnSH exhibit spin-polarized ferromagnetic (FM) behavior, as shown in Table.S1.

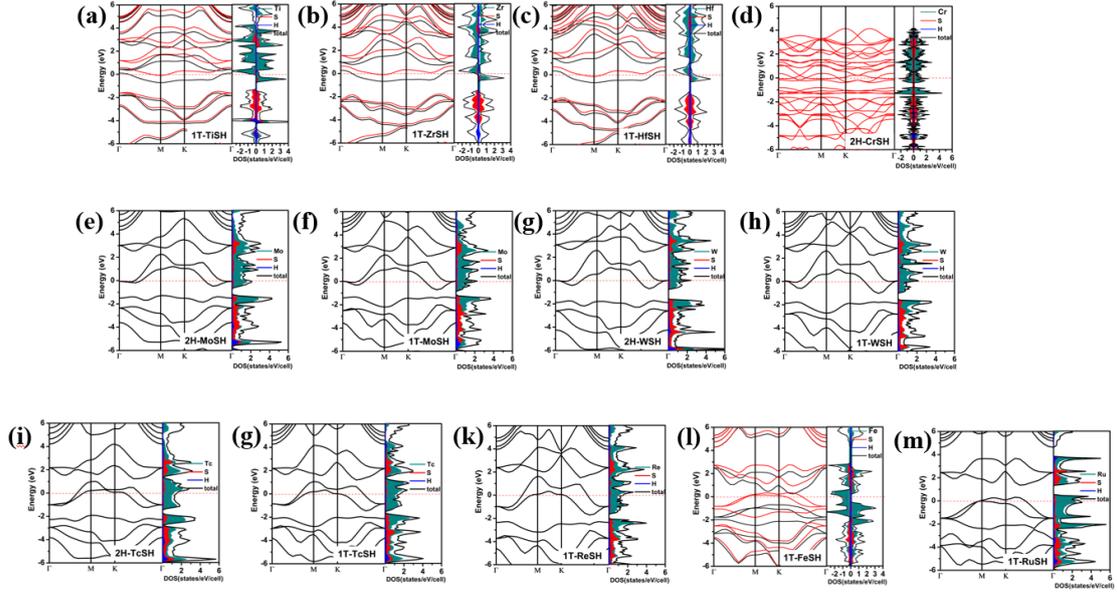

Fig.3 Band structrures and densitity of states of metallic JTMSHs.

In addition to the semiconducting JTMSHs, we have also found many metallic 2H/1T JTMSHs with TM=Ti, Zr, W, etc. (see again Fig.S1 in the supplementary materials). The band structures and density of states of metallic JTMSHs are depicted in Fig.3. It has been observed that JTMSHs with TM=Ti, Zr, Hf, Cr, Mn and Fe exhibit magnetic behaviors, whereas JTMSHs with TM=Mo, W, Tc, Re and Ru are nonmagnetic materials. The calculations of magnetic ground state energy for the magnetic JTMSH materials revealed that 2H-ZrSH, 2H-HfSH and 2H/1T-CrSH are characterized by the spin-polarized AFM ground states, while others show the spin-polarized FM behavior, as shown in Table.S1 and Fig.S1. For all the metallic JTMSHs, the band structures indicate that the DOSs near the Fermi level primarily originate from the $d$-electrons of TM atoms, accompanied by small contributions coming from the $p$-electrons of S atoms. Additionally, for most metallic JTMSHs, the band structures near the Fermi level are rather flat, resulting in the emergence of sharp peaks in the density of states. We note that these sharp peaks might significantly influence the structural stability, as well as the physical and chemical properties of these phases.

## 3.3 Electride states in JTMSHs

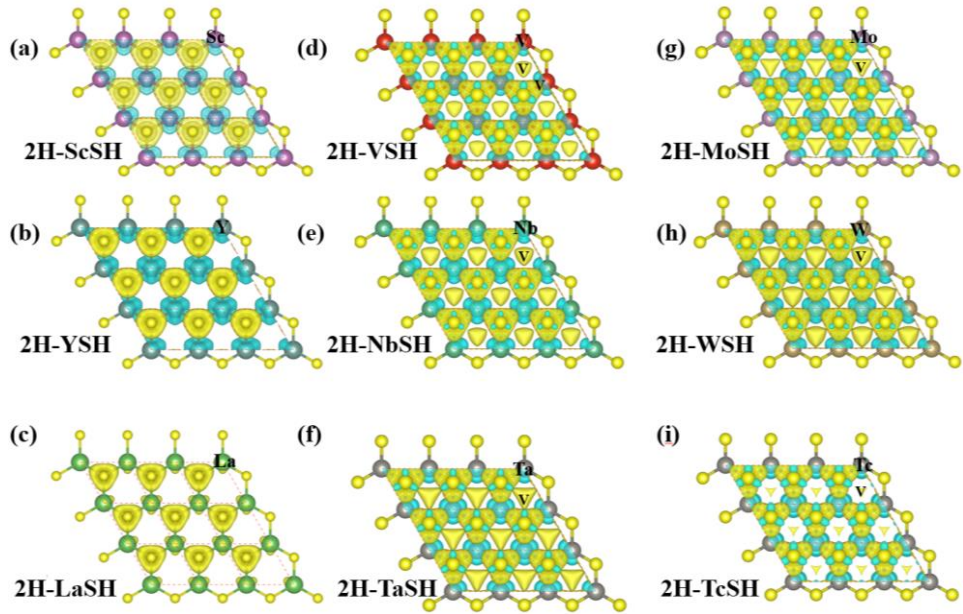

Fig.4 The charge density differences of the non-magnetic 2H-JTMSHs. The yellow atoms refer to sulfur (S) atoms, while the TM atoms are represented by their element symbols displayed in the upper right corner of the structures. The center of electron accumulation in the TM atom layer is represented by the symbol V.

In the next step, we have conducted charge difference density calculations for the selected non-magnetic 2H-TMSHs and observed anomalies in their electronic structure, as depicted in Fig.4. For all the non-magnetic 2H-JTMSHs, it is evident that charges are transferred from TM atoms to S atoms, resulting in the formation of both ionic bonds and polarized covalent bonds with the S atoms. Interestingly, we also noted accumulation of charges in the TM atom layer among the three TM atoms for 2H-TMSHs with TM=V, Nb, Ta, Mo, W, Tc, indicated as the V site in Fig.4. Note that in conventional solids, electrons are typically bound to the atomic nucleus. However, in a specific kind of materials electrons can escape from atomic nucleus and migrate to the interstitial spaces, remaining there and forming the so-called electride states[42]. Because electrons have a significantly smaller mass than anions, such electride materials usually possess distinct characteristics. The mentioned size difference leads to the prominent quantum effect, giving electrides superior conductivity, high electronic mobility, low work function, and high hyperpolarizabilities. Additionally, electrides exhibit electron-rich chemistry and strong electron donation power, making them highly valuable for various applications, including electronics, electrocatalysis, and topological applications[43,44].

The electride states are commonly observed in alkali metals[45] under high pressure, but have also been found in some bulk materials under normal conditions such as $C_{12}A_7$[46], $CaN_2$[47], LaCoSi[48], CeRuSi[44], $Ca_5Pb_3$[49], $AB_2C_3$[50], and $Y(Sc)_xCl_y$[51], but a few 2D materials e.g. $Ca_2N$[52], $LaBr_2$[53] and $M_2C$ with M=Sc, Y[54]. Our findings presented here reveal that 2H-JTMSHs, including experimentally discovered 2H-MoSH, also belong to this category of electride materials. In contrast, no such phenomenon occurs for 1T-JTMSHs (refer to Fig.S3 in the supplementary materials).

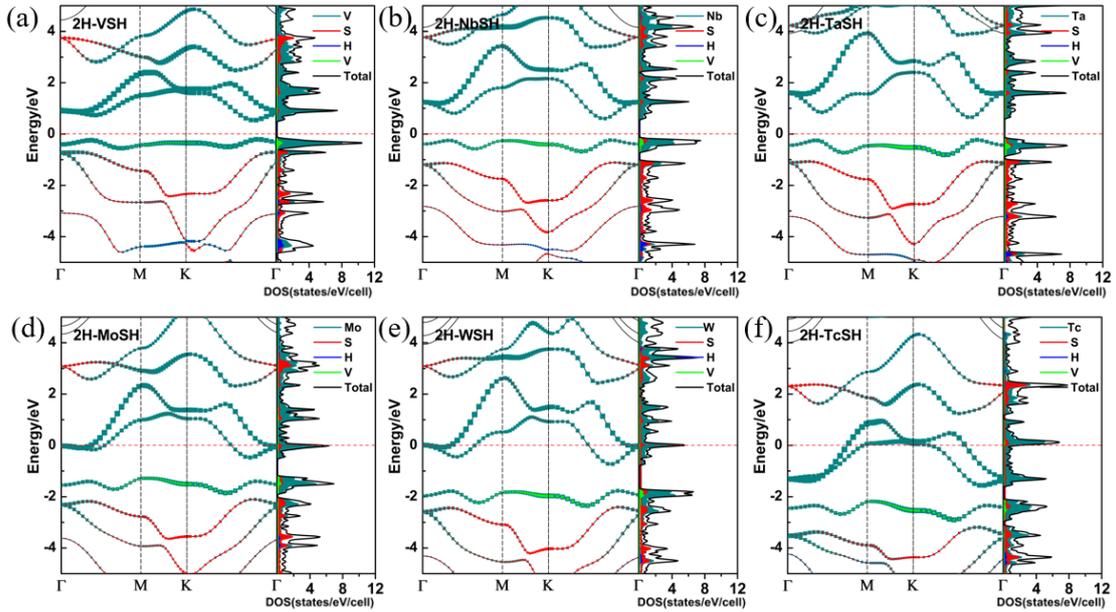

Fig.5 Band structures and DOSs of JTMSHs with contributions of anionic electron states (V) highlighted in green.

To gain deeper understanding of the electride states in JTMSHs, we have calculated the anionic electron bands (V-bands) with contributions coming from the electrons at the V site, as shown in Fig.5. We observe that for the semiconducting JTMSHs with TM=V, Nb and Ta, there are considerable number of anionic electrons in the VBM. However, in the case of the metallic JTMSHs (TM=Mo, W, Tc), the anionic electron states primarily reside below the Fermi level ($E_F$) and are fully occupied, indicating more localized nature of anionic electrons in metallic JTMSHs.

3.4 **Dynamical stabilities of JTMSHs and CDW scenario**

In this work, we have calculated phonon curves of 2H/1T JTMSHs to test their dynamical stability at zero temperature. A phonon curve without negative frequency indicates deep enough potential to prevent structural dissociation. As shown in Fig.S4 in the supplementary materials, no negative frequencies were found in the entire Brillouin zone for the 2H and 1T JTMSHs with M=IIIB and VB subgroup elements, confirming their dynamical stabilities. Similar findings were reported for the metallic JTMSHs. However, it should be noted that some of these materials exhibited phonon dispersion with small negative frequencies in the lowest acoustic branches near the Γ point. Nonetheless, such occurrences are quite common among 2D materials and can be rectified using rotational sum rules[55].

Altought, the negative phonon frequencies are commonly associated with structural instability they can also indicate phase transitions. One such phase transition also may lead to the charge density waves (CDW) scenario. For the CDW materials, when atomic lattice undergoes distortion, the coulomb repulsion energy increases but the electronic band structure energy decreases. If the decrease in energy outweighs the increase in coulomb repulsion energy, a CDW phase transition occurs. For TMDs, many CDW materials has been reported *e.g.* $VTe_2$[56], $ZrTe_2$[57], $TiSe_2$[58], $aSe_2$[59], $NbS_2$[60] and $VS_2$[61].

In context to the above, we note that during phonon curve calculations the Gauss broadening of σ=0.02 Ryd was used. However, by changing this parameter the electron temperature can be varied, allowing for the estimation of its effect on the dynamical stability of materials. In a special case, as the electron temperature increases, the negative frequencies of the phonon curve disappear, resulting in the stabilization of previously unstable materials and occurrence of the CDW phase transition. The application of this method has proven the CDW phase transition in 1T-MoSH in the previous work[27]. By employing the same method, we have found that 1T-MSH with M=Tc, Re, and W as well as 2H-CrSH exhibit CDW properties, as depicted in Fig.5. We notice that the negative frequencies mainly occur at high symmetry M point for 2H-CrSH and 1T-WSH, which is the same as 1T-MoSH in the previous work[27]. In case of other JTMSHs the negative frequencies can be observed at K and M points. Nevertheless, despite using a larger Gauss broadening, the presence of negative frequency persists in the 2H-JTMSHs with TM=Re, Ru, Fe, Os and 1T-OsSH, highlighting inherent instabilities of these structures. This information can be verified in Fig.S5 in the supplementary materials. As we known, the CDW is a complex physical phenomenon. Although we have reported presence of CDW materials in JTMSHs, further experimental and theoretical investigations are expected to gain more indepth understanding of the CDW scenario, including possible mechanisms, phase transition temperature, and resulting phase change products.

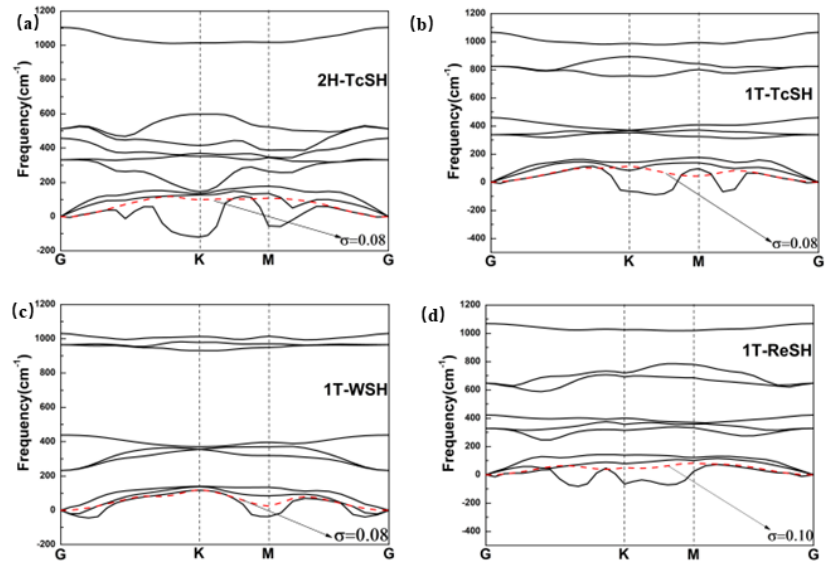

Fig.5 Phonon dispersions of CDW JTMSHs with different Gauss broadenings σ (Ryd). The red lines illustrates that the lowest acoustic branches of the phonon curves become positive with a increasing σ.

### 3.5 Superconducting properties of JTMSHs

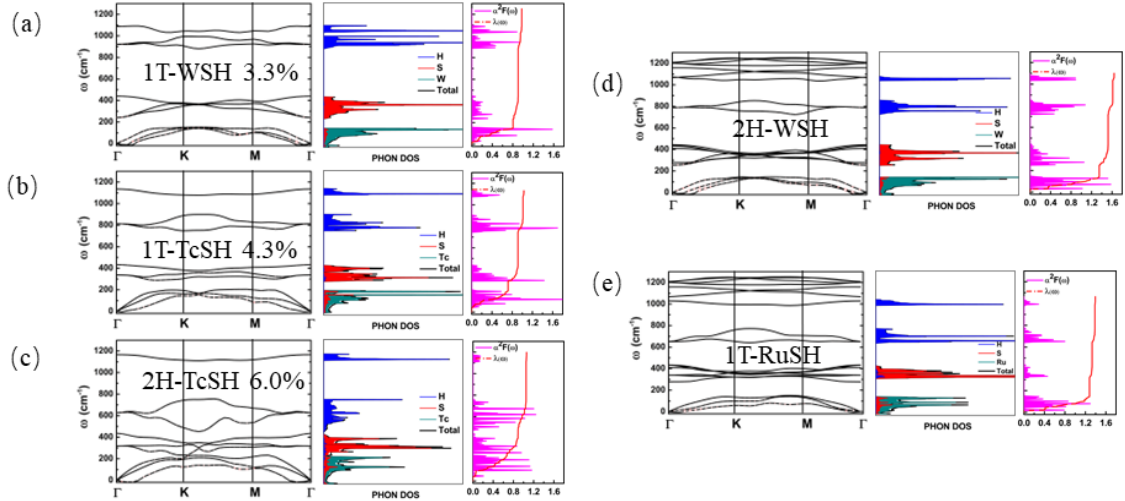

Fig.6 The phonon dispersions, total and partial phonon DOSs as well as $\alpha^2F(\omega)$ function with integrated λ(ω) for various JTMSH monolayer. Hollow red circles in (a)-(e) indicate the phonon linewidth with a radius proportional to its strength. The applied biaxial strains are depicted in each the figure for the CDW materials, namely 1T-WSH, 1T/2H-TcSH.

It is a well-known fact that CDW materials can be engineered via various methods such as the application of external strain. In particular, when subjected to compressive strain, the CDW phase of both 1T-WSH and 1T/2H TcSH can be completely suppressed. Fig.6 a-c illustrate the phonon curves of these materials under compressive biaxial strains. Apart from minor errors around the Γ points, due to the 2D nature of these materials, there are no negative frequencies in the entire Brillouin zone. This indicates that such materials are dynamically stable under strain. Importantly, the previous work on 1T-MoSH also demonstrated similar suppression of the CDW phase[27].

Furthermore, that strain can lead to the potential superconductivity in the discussed 2D CDW materials. As show in Table 1, when subjected to a compressive strain within 10%, the $T_C$ of 1T-WSH, 1T-TcSH and 2H-TcSH can reach maximum value of 13.8, 16.2 and 24.2 K respectively. Moreover, the $T_C$ of 1T-WSH and 1T-TcSH decrease with increasing strain, while the $T_C$ of 2H-TcSH increases. We also found stable intrinsic superconductors of 2H-WSH and 1T-RuSH. The $T_C$ of 2H-WSH and 1T-RuSH can reach 17.0 and 8 K, respectively. Fig.6 shows $\alpha^2F(\omega)$ and λ(ω) as well as the phonon dispersions and projected phonon DOSs for all superconducting JTMSHs. From the total and partial phonon DOSs in Figure 6, it is apparent that all the phonon curves exhibit similar characteristics. For JTMSHs, significant fraction of total λ comes from the low-frequency vibrational modes, which are dominated by the heavier TM atoms. That said, superconductivity in JTMSHs appears to primarily stem from the TM atoms, with a minor contribution from the S atoms.

Table.1 Dependence of $\lambda$, $\omega$ and $T_C$ on the in-plane biaxial strain of the JTMSHs

| Material | Biaxial Strain | λ | ω | Tc/K |
|---|---|---|---|---|
| 1T-WSH | 3.3% | 1.11 | 170.6 | 13.8 |
|  | 5.6% | 0.57 | 290.7 | 5.8 |
| 1T-TcSH | 4.3% | 1.02 | 225.3 | 16.2 |
|  | 7.0% | 0.65 | 282.4 | 8.23 |
| 2H-TcSH | 6.0% | 1.00 | 287.9 | 20.1 |
|  | 10.0% | 2.48 | 147.3 | 24.2 |
| 2H-WSH | - | 1.68 | 135.1 | 17.03 |
| 2H-RuSH | - | 1.27 | 87.6 | 8.43 |

## 4. Summary and Conclusion

In summary, we have conducted extensive first-principles calculations to explore the possible two-dimensional structures of 2H and 1T Janus transition-metal dichalcogenide monolayers. Their synthesis potential was determined based on the detailed analysis of the thermodynamic and dynamic stabilities of these materials. In particular, our simulations revealed that 1T-JTMSHs are generally more stable than the 2H-JTMSHs with the same TM element except for LaSH. Moreover, it was shown that the 1T/2H-JTMSHs with M=IIIB and VB subgroup elements are nonmagnetic indirect-gap semiconductors, while 2H-JTMSHs with M=Ti, Zr, Hf, and 1T-JTMSHs with M=Cr, Mn are small band gap magnetic semiconductors. On the other hand, the 1T JTMSHs with TM=Ti, Zr, Hf, Cr, Fe and 2H-CrSH were found to be magnetic metals, whereas 1T-JTMSHs with TM=W, Tc, Re, Ru and 2H-JTMSHs with TM=Tc, W are nonmagnetic metals. More intriguingly, however, we have observed anomalies in the electronic structures of some 2H-JTMSHs with TM=V, Nb, Ta, Mo, W, Tc. Specifically, charge was found to accumulate in the TM atom layer, suggesting formation of electride states. Next, comprehensive phonon calculations were performed, proving existence of CDW scenario in 1T-JMSHs with M=Tc, Re, W and 2H-TcSH materials. In addition, we have demonstrated that the CDW in 1T-WSH and 1T/2H TcSH can be suppressed under compressive strain. Finally, the applied compressive strain was found to be responsible for induction of phonon-mediated superconducting state. In details, under compressive strain within 10%, the $T_C$ in 1T-WSH, 1T-TcSH, and 2H-TcSH was shown to reach values of 13.8, 16.2, and 24.2 K, respectively. Furthermore, we have discovered intrinsic superconductivity in 2H-WSH and 1T-RuSH, with $T_C$ values of 17.0 K and 8 K, respectively.

In conclusion, our results demonstrate that the family of 2D JTMSHs presents rich physical and chemical properties, significantly extending the potential applications of 2D materials based on transition metals. While predicted semiconducting phases possess in-direct band gaps that are larger than their counterparts in parent TMDs[17], (meaning they are not energetically favorable for some applications *e.g.* in terms of efficient light emitters), they may be still of use for low-dimensional integrated circuits and high-speed electronics. Still, the suggested metallic phases appear to be even more interesting, mainly due to the possible occurrence of conventional superconductivity. In the first place, these phases are expected to yield relatively high $T_C$ values, which are two (or in the case of compressive strain, even three times) higher than the transition temperature observed for the exemplary 2D superconductor, $LiC_6$[62,63]. Secondly, by inspecting in detail the corresponding band structures of superconducting JTMSHs, it becomes apparent that these materials are shallow conduction band materials, indicating that their superconducting properties may be notably influenced by the non-adiabatic effects[64]. In fact, the situation may be similar to what has been observed in bulk materials such as bismuthates[65] or the already mentioned 2D $LiC_6$[63]. Hence, despite conducted here analysis there are still many unresolved issues that require further theoretical and experimental investigations in terms of 2D JTMSHs. This includes not only role of the non-adiabatic effects in shaping related superconducting phases but also the phase transition mechanism or the phase transition products of

CDW scenario.

**CRediT authorship contribution statement**

DaWei Zhou: Conceptualization, Investigation, Writing – original draft. Dominik Szczęśniak: Conceptualization, Supervision, Writing – review & editing. Pan Zhang: Investigation. Zhuo Wang: Investigation. Chunying Pu: Conceptualization, Investigation, Supervision, Writing – review & editing.

**Declaration of competing interest**

The authors declare that they have no known competing financial interests or personal relationships that could have appeared to influence the work reported in this paper

**Acknowledgement**

This research was funded by the National Natural Science Foundation of China (grant No. 52272219, U1904612), the Natural Science Foundation of Henan Province (grant Nos. 222300420506.)

**Appendix A. Supplementary material**

The total density of states of JTMSHs including 1T and 2H phases with the spin-polarization effects. The calculated magnetic ground state energy of the ferromagnetic (FM), antiferromagnetic (AFM), and nonmagnetic (NM) of TMSH materials using a (2×2) supercell. The band structures and total density of states for the magnetic semiconducting JTMSHs. The charge density differences of the non-magnetic 1T-JTMSHs. Phonon curves of stable and unstable JTMSHs.